\documentclass[draft]{agujournal2019}
\usepackage{url} %this package should fix any errors with URLs in refs.
\usepackage{lineno}
\usepackage[inline]{trackchanges} %for better track changes. finalnew option will compile document with changes incorporated.
\usepackage{soul}
\usepackage{multirow}
\usepackage{colortbl}
\usepackage{amsmath,siunitx}
\usepackage[version=4]{mhchem}
\usepackage{booktabs} % To thicken table lines
\usepackage{tikz}
\usetikzlibrary{arrows.meta}
\draftfalse

\journalname{JGR: Atmospheres}

\begin{document}

%%%%%%%%%%%%%%%%%%%%%%%%%%%%%%%%%%%%%%%%%%%%%%%
%  TITLE
%
% (A title should be specific, informative, and brief. Use
% abbreviations only if they are defined in the abstract. Titles that
% start with general keywords then specific terms are optimized in
% searches)
%
%%%%%%%%%%%%%%%%%%%%%%%%%%%%%%%%%%%%%%%%%%%%%%%

\title{Radiative forcing by super-volcano eruptions}

%%%%%%%%%%%%%%%%%%%%%%%%%%%%%%%%%%%%%%%%%%%%%%%
%
%  AUTHORS AND AFFILIATIONS
%
%%%%%%%%%%%%%%%%%%%%%%%%%%%%%%%%%%%%%%%%%%%%%%%

\authors{Eirik R. Enger\affil{1}, Rune Graversen\affil{1}, Audun Theodorsen\affil{1}}

\affiliation{1}{UiT The Arctic University of Norway, Tromsø, Norway}

\correspondingauthor{Eirik R. Enger}{eirik.r.enger@uit.no}

\begin{keypoints}
  \item The linear RF dependence on AOD breaks down for eruptions larger than Mt. Pinatubo
  \item The RF to AOD ratio has a time-after-eruption dependence on eruption latitude
  \item Temperature and RF peak values has a linear dependence and reaches an upper limit
\end{keypoints}

\begin{abstract}
  We investigate the climatic effects of volcanic eruptions spanning from Mt.\
  Pinatubo-sized events to super-volcanoes. The study is based on ensemble simulations
  in the Community Earth System Model Version 2 (CESM2) climate model using the Whole
  Atmosphere Community Climate Model Version 6 (WACCM6) atmosphere model. Our analysis
  focuses on the impact of different \ce{SO2}-amount injections on stratospheric aerosol
  optical depth (AOD), effective radiative forcing (RF), and global temperature
  anomalies. Unlike the traditional linear models used for smaller eruptions, our
  results reveal a non-linear relationship between RF and AOD for larger eruptions. We
  also uncover a notable time-dependent decrease in aerosol forcing efficiency across
  all eruption magnitudes during the first post-eruption year. In addition, the study
  reveals that larger as compared to medium-sized eruption events produce a delayed and
  sharper peak in AOD, and a longer-lasting temperature response while the time
  evolution of RF remains similar between the two eruption types. When including the
  results of previous studies, we find that relating \ce{SO2} to any other parameter is
  inconsistent across models compared to the relationships between AOD, RF, and
  temperature anomaly. Thus, we expect the largest uncertainty in model codes to relate
  to the chemistry and physics of \ce{SO2} evolution. Finally, we find that the peak RF
  approaches a limiting value, and that the peak temperature response follows linearly,
  effectively bounding the temperature anomaly to at most \(\sim\SI{-12}{\kelvin}\).
\end{abstract}

% Atmospheres now.
% https://www.agu.org/Publish-with-AGU/Publish/Author-Resources/Text-requirements
\section*{Plain Language Summary}

% Here are instructions on writing a Plain Language Summary:
% https://www.agu.org/Share-and-Advocate/Share/Community/Plain-language-summary

Volcanic eruptions can have a significant impact on the Earth's climate. Eruptions large
enough that the aerosols they emit reach the stratosphere cause a cooling effect by
reflecting sunlight. Typically, an eruption is measured by its impact on the opacity of
the stratosphere and the change in the energy balance at the top of the atmosphere. The
two measures are often assumed to be linearly related, but the linearity is tested only
against eruptions seen in the last two millennia. We use a coupled climate model to
simulate the impact of eruptions of sizes up to the largest known eruptions. The
smallest eruptions we simulate are still large enough to cause global climate effects.
We find a clear non-linear relationship for eruptions larger than the ones seen in the
past two millennia. Our simulations and supporting data shows that the eruption latitude
significantly influences the development of the relationship between energy imbalance
and stratospheric opacity with time after the eruption. Additionally, we find evidence
that the peak energy imbalance reaches a limit, and that the peak temperature response
follows linearly with the peak energy imbalance, also reaching a limiting value.

% FIXME: related to this, there is an index that can be added during publishing. Find
% the list here:
% https://www.agu.org/publish-with-agu/publish/author-resources/index-terms
% 8409 - Atmospheric effects (0370)
% 8408 - Volcano/climate interactions (1605, 3309, 4321)
% 1627 - Coupled models of the climate system
% 0305 - Aerosols and particles (0345, 4801, 4906)
% 3363 - Stratospheric dynamics
% 0370 - Volcanic effects

\section{Introduction}

% NOTE: Suggested layout for the introduction
% - The objectives of the work.
% - The justification for these objectives: Why is the work important?
% - Background: Who else has done what? How? What have we done previously?
% - Guidance to the reader: What should the reader watch for in the paper? What are the
%   interesting high points? What strategy did we use?
% - Summary/conclusion: What should the reader expect as conclusion? In advanced
%   versions of the outline, you should also include all the sections that will go in
%   the Experimental section (at the level of paragraph subheadings) and indicate what
%   information will go.

Effective radiative forcing (RF) and stratospheric aerosol optical depth (AOD) are
crucial metrics representing the energy imbalance at top-of-the-atmosphere (TOA) and the
stratospheric opacity due to aerosol scattering, respectively. They are extensively used
to quantify the impact of major volcanic eruptions. The assumption of a linear
dependency of RF on AOD is commonly adopted \cite{myhre2013,andersson2015}, and applying
such a linear relationship has yielded reasonably accurate estimates in climate model
simulations of volcanic eruptions
\cite{mills2017,hansen2005,gregory2016,marshall2020,pitari2016b}. Yet, a wide spread in
the estimated aerosol forcing efficiencies (RF normalised by AOD) exists among studies,
spanning approximately from \(\sim \SI{-15}{\watt\metre^{-2}\ce{AOD}^{-1}}\)
\cite{pitari2016b} to \(\sim \SI{-25}{\watt\metre^{-2}\ce{AOD}^{-1}}\) \cite{myhre2013}.
Additionally, these estimates are predominantly based on small eruptions with AOD values
up to at most \(\sim 0.7\).

Although \ce{H2O}, \ce{N2}, and \ce{CO2} are the most abundant gases emitted by
volcanoes \cite{robock2000}, sulphur species such as \ce{SO2} provide a greater
influence due to the comparatively high background concentrations of the former gases in
the atmosphere. The transformation of \ce{SO2} molecules through reactions with \ce{OH}
and \ce{H2O} leads to the formation of sulphuric acid (\ce{H2SO4}) \cite{robock2000},
which scatters sunlight thereby elevating planetary albedo and reducing the RF. As the
conversion from \ce{SO2} to \ce{H2SO4} occurs over weeks \cite{robock2000}, the peak RF
experiences a slight delay from the eruption's peak \ce{SO2} injection. The lifetime of
the \ce{H2SO4} aerosols in the stratosphere depends on various factors, including
latitude \cite{marshall2019, toohey2019}, volcanic plume height \cite{marshall2019},
aerosol size \cite{marshall2019}, the quasi-biennial oscillation phase
\cite{pitari2016b} and the season of the year (determining to which hemisphere aerosols
are transported) \cite{toohey2011,toohey2019}. In the case of tropical eruptions,
aerosols are typically transported poleward in the stratosphere and descend back to
mid-latitude troposphere within one to two years \cite{robock2000}. Upon descending
below the tropopause, these aerosols are readily removed by wet deposition
\cite{liu2012}.

Before the current era of significant anthropogenic climate forcing, volcanic eruptions
were the primary forcing mechanism dictating Earth's climate variability during the
Holocene period \cite{sigl2022}. Despite this substantial impact, few climate-model
experiments have included volcanic forcing when simulating climate evolution during the
Holocene \cite{sigl2022}, likely implying an exaggerated positive forcing
\cite{gregory2016,solomon2011}. This absence of persistent cooling is one of several
factors that have been suggested to contribute to the common disparity between simulated
and observed global warming \cite{andersson2015}. Despite extensive attention on
understanding the way volcanic eruptions influence climate, questions regarding aerosol
particle processes---such as growth and creation rates when \ce{OH} is scarce---remain
unanswered \cite<e.g.~>[]{robock2000,zanchettin2019,marshall2020,marshall2022}. These
processes impact aerosol scattering efficiency and potentially the RF to AOD
relationship. \citeA{marshall2020} observe higher aerosol forcing efficiency in
post-eruption years \(2\) and \(3\) compared to year 1, and attribute this post-eruption
increase in aerosol forcing efficiency to strong spatial concentration in the initial
year and subsequent distribution of aerosols over a larger area. This spatial
redistribution increases the albedo per global mean AOD thereby causing a stronger RF to
AOD ratio \cite{marshall2020}.

Previous studies of both Mt.\ Pinatubo \cite{mills2017,hansen2005} and volcanoes within
the instrumental era \cite{gregory2016} have been used to estimate the relationship
between the RF energy imbalance and change in AOD caused by volcanic eruptions. While
\citeA{myhre2013} employ a formula scaling RF by AOD to obtain
\(\SI{-25}{\watt\metre^{-2}\mathrm{AOD}^{-1}}\), recent literature reports estimates
down to \(\SI{-19.0(5)}{\watt\metre^{-2}\mathrm{AOD}^{-1}}\) \cite{gregory2016} and
\(\SI{-18.3(10)}{\watt\metre^{-2}\mathrm{AOD}^{-1}}\) \cite{mills2017}. Synthetic
volcano simulations in \citeA{marshall2020} yield a scaling factor of
\(\SI{-20.5(2)}{\watt\metre^{-2}\mathrm{AOD}^{-1}}\) across an ensemble of \(82\)
simulations featuring varying injection heights and latitudes of volcanic emissions,
with injected \ce{SO2} ranging from \(10\) to \(\SI{100}{\tera\gram(\ce{SO2})}\).

A similar simulation setup, albeit with notable differences, was conducted by
\citeA{niemeier2015}, involving an ensemble of \(14\) levels of injected sulphur
spanning between \(\SI{1}{\tera\gram(\ce{S})\mathrm{yr}^{-1}}\)
(\(\SI{2}{\tera\gram(\ce{SO2})\mathrm{yr}^{-1}}\)) and
\(\SI{100}{\tera\gram(\ce{S})\mathrm{yr}^{-1}}\)
(\(\SI{200}{\tera\gram(\ce{SO2})\mathrm{yr}^{-1}}\)). These geoengineering simulations
maintained continuous sulphur injections, running until a steady sulphur level was
achieved. Results indicated an inverse exponential relationship between RF and injected
\ce{SO2} rate, converging to \(\SI{-65}{\watt\metre^{-2}}\)
(Eq.~\ref{eq:niemeier_exponential}). Even the \(100\times\) Mt.\ Pinatubo super-volcano
simulation by \citeA{jones2005}, which obtained a peak RF of
\(\SI{-60}{\watt\metre^{-2}}\), is below the suggested limit of
\(\SI{-65}{\watt\metre^{-2}}\). Moreover, \citeA{timmreck2010} find a peak RF anomaly of
\(\SI{-18}{\watt\metre^{-2}}\) from a \(\SI{1700}{\tera\gram(\ce{SO2})}\) eruption
simulation, which corresponds well with the function estimated by \citeA{niemeier2015}
at the given \ce{SO2} level. Several studies have demonstrated a linear relationship of
approximately \(-\SI{20}{\watt\metre^{-2}\mathrm{AOD}^{-1}}\) between RF and AOD,
although substantial variability exists in the slope among studies
\cite{mills2017,hansen2005,gregory2016,marshall2020,pitari2016b}. Moreover, a
time-after-eruption dependence on the RF to AOD ratio is found in \citeA{marshall2020},
whereas \citeA{niemeier2015} revealed a non-linear relationship between RF and injected
\ce{SO2} rate. Thus, a consensus on the relationship between injected \ce{SO2}, AOD, and
RF has yet to be established.

One avenue that has garnered considerable attention is comparing the magnitude of
volcanic or volcano-like forcings to increased \ce{CO2} levels. Several studies explore
the connection between volcanic forcing and the climate sensitivity to a doubling of
\ce{CO2}
\cite{boer2007,marvel2016,merlis2014,ollila2016,richardson2019,salvi2022,wigley2005}.
The comparison of forcing from volcanoes and \ce{CO2} aims to mitigate the large
uncertainty in estimates of the sensitivity of the real climate system. Inferring
climate sensitivity from volcanic eruption events has been attempted as a way to
constrain the sensitivity \cite{boer2007} by assuming that volcanic and \ce{CO2}
forcings produce similar feedbacks \cite{pauling2023}. Earlier studies suggest the
potential for constraining equilibrium cilmate sensitivity (ECS) using volcanoes
\cite{bender2010}, provided that ECS is constrained by effective radiative forcing (ERF)
rather than instantaneous radiative forcing (IRF), as ERF accounts for rapid atmospheric
adjustments in contrast to IRF \cite{richardson2019}. However, other studies refute this
approach, pointing out that different sensitivities of volcanic forcing and \ce{CO2}
doubling seem to exist \cite{douglass2006}, or that constraining the ECS by ERF lacks
accuracy due to the precision of climate simulations \cite{boer2007,salvi2022}. Although
ERF offers a more suitable indicator of forcing than IRF
\cite{marvel2016,richardson2019}, more recent studies conclude that ECS cannot be
constrained from volcanic eruption events \cite{pauling2023}.

Employing eruptions in the medium to super-volcano size enhances the signal-to-noise
ratio without necessitating an extensive and computationally expensive ensemble, and as
such, is a tempting way to mimic a large ensemble of smaller volcanic eruptions.
However, the AOD, RF, and temperature signatures are not necessarily a simple scaling of
that of smaller volcanic eruptions. Previous studies have simulated super-volcanoes
using AOD as the input forcing, where the AOD was that of Mt.\ Pinatubo scaled by a
factor of one hundred \cite{jones2005}. This approach may yield incorrect results, both
because the peak of the AOD may be too small or too big, but also because the evolution
of the AOD could be inappropriate. Likewise, a different AOD evolution may be found from
similar size eruptions, but at different latitudes. To investigate this issue, our
simulations are based on four levels of injected \ce{SO2} covering three orders of
magnitude and the inclusion of one high latitude eruption of the second largest injected
\ce{SO2} case.

We conducted ensemble simulations of volcanic eruptions in the Community Earth System
Model Version 2 (CESM2) coupled with the Whole Atmosphere Community Climate Model
Version 6 (WACCM6). The ensembles span four different levels of injected \ce{SO2}:
\(\SI{26}{\tera\gram(\ce{SO2})}\), \(\SI{400}{\tera\gram(\ce{SO2})}\),
\(\SI{1629}{\tera\gram(\ce{SO2})}\) and \(\SI{3000}{\tera\gram(\ce{SO2})}\). Details
regarding the experimental setup are provided in section~\ref{sec:method}. Our findings
reveal non-linear RF to AOD dependencies for medium to super-volcano size eruptions.
Additionally, we observe a time-dependent variation in the RF to AOD ratio, detailed in
section~\ref{sec:results} and discussed in section~\ref{sec:discussion}. Furthermore,
our data, along with insights from previous studies, suggest that the RF dependency on
injected \ce{SO2} identified by \citeA{niemeier2015} acts as a lower boundary. Our
conclusions are presented in section~\ref{sec:conclusions}.

\section{Method}\label{sec:method}

\subsection{Model}

We use the CESM2 \cite{danabasoglu2020} in conjunction with the WACCM6
\cite{gettleman2019} and the fully dynamical ocean component Parallel Ocean Program
version 2 (POP2) \cite{smith2010, danabasoglu2020}. The atmosphere model was run at a
nominal \(\SI{2}{\degree}\) resolution with \(70\) vertical levels in the middle
atmosphere (MA) configuration.

The WACCM6 version employed in the MA configuration uses the three mode version of the
Modal Aerosol Module (MAM3) \cite{gettleman2019}, a simplified and computationally
efficient default setting within the Community Atmosphere Model version 5 (CAM5)
\cite{liu2016}, as described in \citeA{liu2012}. The MAM3 was developed from MAM7 and
features the modes Aitken, accumulation, and coarse \cite{liu2016}.

\subsection{Simulations}

Appendix A provides a description of the simulation setup and utilised output variables.
Table~\ref{tab:simulation-overview} summarises the simulations, encompassing four
\ce{SO2} injection magnitudes and up to four seasons: 15 February, 15 May, 15 August,
and 15 November. The magnitudes vary over three orders of magnitude:
\(\SI{26}{\tera\gram(\ce{SO2})}\), \(\SI{400}{\tera\gram(\ce{SO2})}\),
\(\SI{1629}{\tera\gram(\ce{SO2})}\), and \(\SI{3000}{\tera\gram(\ce{SO2})}\).

The smallest eruption case, C2W\(\downarrow\), is similar in magnitude as compared to
events like Mt.\ Pinatubo
\cite<\(\sim10\)--\(\SI{20}{\tera\gram(\ce{SO2})}\);>[]{timmreck2018} and Mt.\ Tambora
\cite<\(\sim\SI{56.2}{\tera\gram(\ce{SO2})}\);>[]{zanchettin2016}. The intermediate
case, C2W\(-\), resembles the magnitude of the Samalas eruption in 1257
\cite<\(\sim{144}\)--\(\SI{170}{\tera\gram(\ce{SO2})}\);>[]{vidal2016}, while the second
largest and largest eruption cases, C2W\(\uparrow\) and C2W\(\uparrow\uparrow\), is in
the likely range of the Young Toba Tuff (YTT) eruption occurring about
\(\SI{72000}{\mathrm{yr}}\) ago
\cite<\(100\)--\(\SI{10000}{\tera\gram(\ce{SO2})}\);>[]{jones2005}. All eruptions were
situated at the equator (\(\SI{0}{\degree N}\), \(\SI{1}{\degree E}\)) with \ce{SO2}
injected between \(\SI{18}{\kilo\meter}\) and \(\SI{20}{\kilo\meter}\) altitude.
Collectively, the four tropical eruption cases C2W\(\downarrow\), C2W\(-\),
C2W\(\uparrow\), and C2W\(\uparrow\uparrow\) are referred to as C2WTrop. An additional
high-latitude eruption ensemble, labelled C2WN\(\uparrow\), of the same injected
\ce{SO2} magnitude as C2W\(\uparrow\) was simulated at \(\SI{56}{\degree N}\),
\(\SI{287.7}{\degree E}\) with a six-month separation (15 February and 15 August).

\begin{table*}
  \centering

  \caption{Simulations done with the CESM2\(^{a}\)}\label{tab:simulation-overview}%
  \begin{center}
    \begin{tabular}[c]{cccccc}
      \toprule
      Ensemble name                   & \(\si{\tera\gram(\ce{SO2})}\)         &
      Lat [\si{\degree\mathrm{N}}]    & Lon [\si{\degree\mathrm{E}}]          & Alt [\si{\kilo\metre}] & Eruption months \\
      \midrule
      C2W\(\uparrow\uparrow\)         & \(3000\)                              &
      \(\hphantom{1}0\)               & \(\hphantom{28}1\hphantom{.7}\)       &
      \(18\)--\(20\)                  & \hphantom{Feb,}May,\hphantom{Aug,}Nov                                            \\
      C2WN\(\uparrow\)                & \(1629\)                              &
      \(56\)                          & \(287.7\)                             &
      \(18\)--\(20\)                  & Feb,\hphantom{May,}Aug\hphantom{,Nov}                                            \\
      C2W\(\uparrow\)                 & \(1629\)                              &
      \(\hphantom{1}0\)               & \(\hphantom{28}1\hphantom{.7}\)       & \(18\)--\(20\)
                                      & Feb,May,Aug,Nov                                                                  \\
      C2W\(-\)                        & \(\hphantom{1}400\)                   &
      \(\hphantom{1}0\)               &
      \(\hphantom{28}1\hphantom{.7}\) &
      \(18\)--\(20\)                  & Feb,May,Aug,Nov                                                                  \\
      C2W\(\downarrow\)               & \(\hphantom{14}26\)                   &
      \(\hphantom{1}0\)               &
      \(\hphantom{28}1\hphantom{.7}\) & \(18\)--\(20\)
                                      &
      Feb,May,Aug,Nov                                                                                                    \\
      \toprule
      \multicolumn{6}{l}{\parbox{\linewidth}{\(^{a}\)The ensembles C2WN\(\uparrow\) and C2W\(\uparrow\) have the same
          eruption magnitude, but while C2W\(\uparrow\) is located at the equator, C2WN\(\uparrow\) is
          located at a high northern latitude. C2W\(\uparrow\uparrow\), C2W\(-\) and C2W\(\downarrow\) are located
          at the equator, but with different magnitudes compared to C2W\(\uparrow\). The three smallest
          tropical ensembles have four members, indicated by the number of eruption months, while
          the northern latitude and the extra large super-volcano ensemble consists of two
          members.}}
    \end{tabular}
  \end{center}
\end{table*}

\section{Results}\label{sec:results}

% NOTE: the results should be laid out in a logical way, with the most
% interesting/important stuff first, then tangents that dig deeper at specific things
% later.
% 1. RF to AOD time-after-eruption dependence should be top priority (8 figs atm.)
% 2. Then probably temperature scaling since we discuss the shape of both AOD and RF
%    time series before that (MOTIVATION: can we expect a specific temperature time
%    series shape based on the shape of either of or both of the RF and AOD time
%    series?)
% 3. If there is something interesting to say about the rest of the figures (all the
%    comparing of parameters), then this should come here.

\subsection{Analysis of the time series}

\begin{figure}
  \centering
  \includegraphics{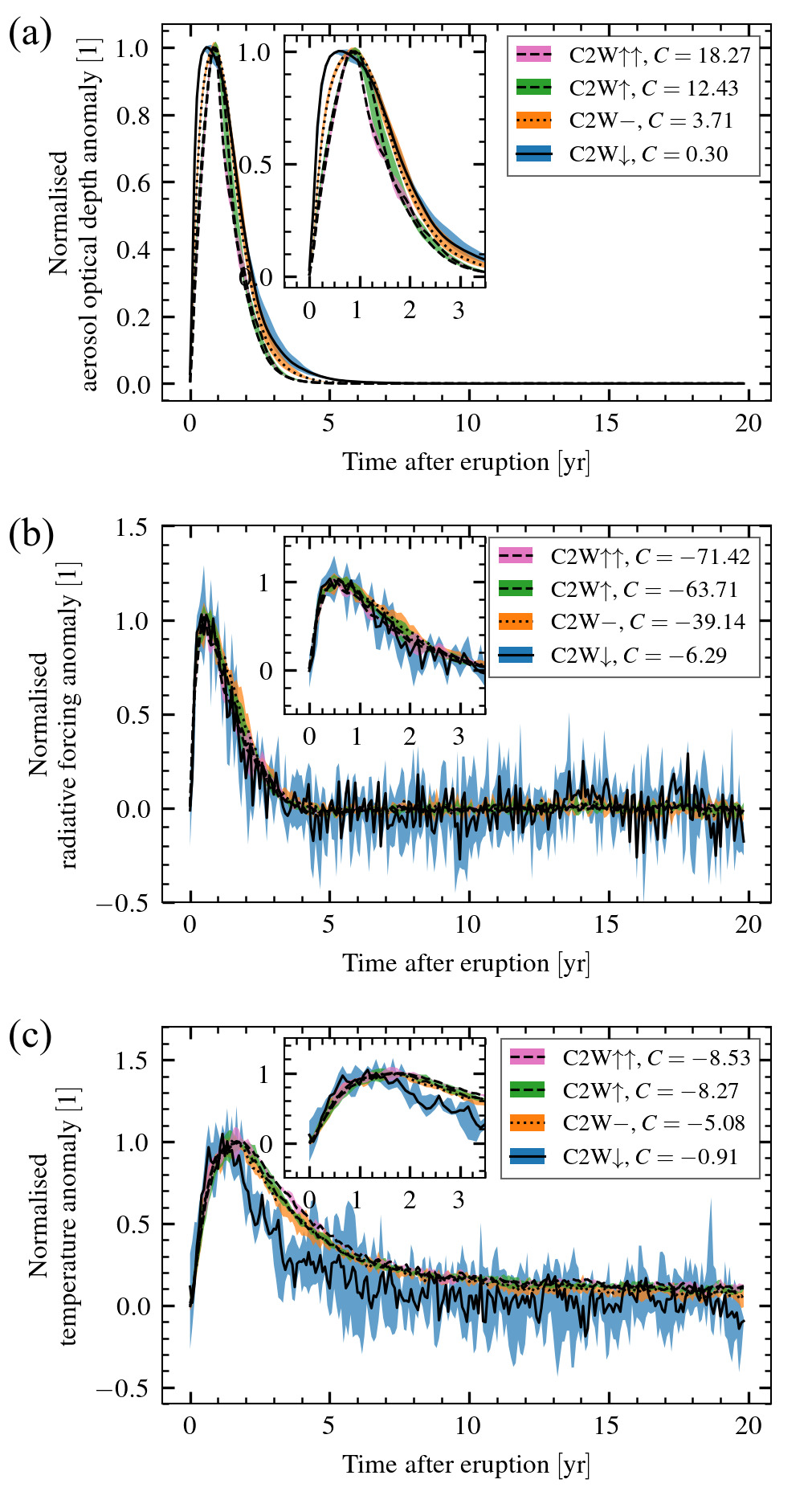}

  \caption{AOD (a), RF (b) and temperature response (c) time series to the four tropical
    volcanic eruption cases, C2W\(\downarrow\), C2W\(-\), C2W\(\uparrow\), and
    C2W\(\uparrow\uparrow\). The time series have been normalised to have peak values at
    unity, where \(C\) is the normalisation constant. Black lines indicate the median across
    the ensembles, while shading marks the 5th and 95th
    percentiles.}\label{fig:1_compare_waveform}%
\end{figure}

Figure~\ref{fig:1_compare_waveform} presents time series of global mean AOD, RF, and
surface air temperature. The black lines represent the medians across the ensembles,
while shading indicates the 5th to 95th percentiles. The four distinct forcing
magnitudes (C2W\(\downarrow\), C2W\(-\), C2W\(\uparrow\), and C2W\(\uparrow\uparrow\))
outlined in table~\ref{tab:simulation-overview} have been used. The time series in
Fig.~\ref{fig:1_compare_waveform} are normalised by setting the peak value to unity,
defined based on the peak of a fit from a Savitzky-Golay filter of 3rd order and a
one-year window length \cite{savitzky1964}.

A notable feature across the subfigures of Fig.~\ref{fig:1_compare_waveform} is the peak
occurrence of the C2W\(\downarrow\) case compared to the larger eruption cases. The peak
of C2W\(\downarrow\) arrives earlier for both AOD (Fig.~\ref{fig:1_compare_waveform}a)
and temperature (Fig.~\ref{fig:1_compare_waveform}c), while the RF time series in
Fig.~\ref{fig:1_compare_waveform}b are all indistinguishable. Cases C2W\(-\),
C2W\(\uparrow\), and C2W\(\uparrow\uparrow\) are indistinguishable in their temperature
development, and while C2W\(\downarrow\) peaks at an earlier time, it decays similarly
to the other cases. Interestingly, the same development between C2W\(-\) and
C2W\(\uparrow\) is not found in the AOD time series. C2W\(\downarrow\) peaks at an
earlier time, but also spends more time around the peak and as such decays at a later
time post-eruption. Likewise, C2W\(-\) has a faster rise and slower decay compared to
C2W\(\uparrow\), but where both peak at a similar time. C2W\(\uparrow\) and
C2W\(\uparrow\uparrow\) have similar AOD developments, but where C2W\(\uparrow\uparrow\)
show a slightly faster decay from the peak.

The timescale of the perturbation of AOD and RF is shorter than that of the temperature.
While the AOD and RF time series return to their equilibrium state within roughly three
years, the temperature time series remain heavily perturbed three years post-eruption.
Even when running the simulations for 20 years post-eruption, the temperature time
series are still decaying.

\subsection{RF dependency on AOD}

\begin{figure}
  \centering
  \includegraphics{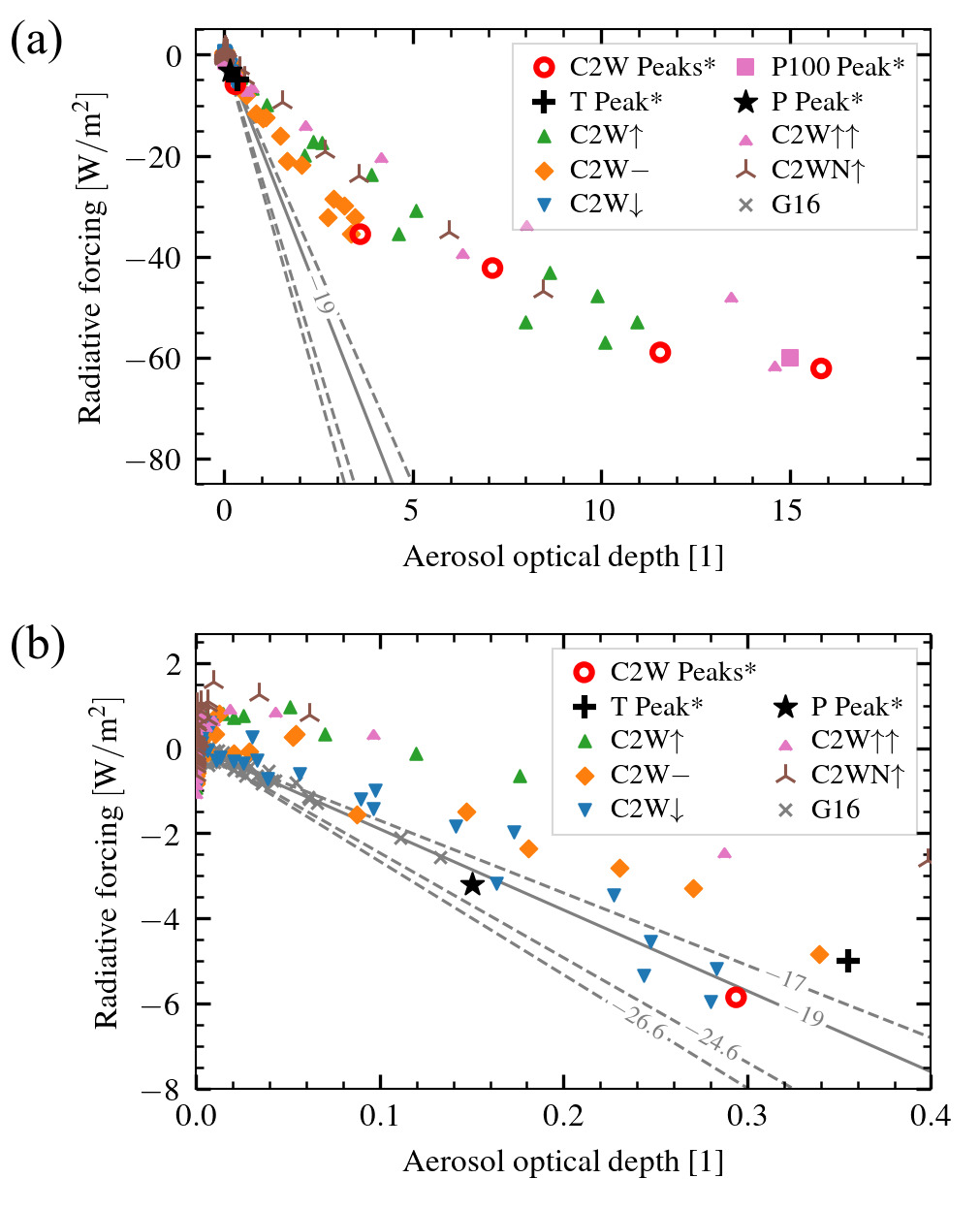}

  \caption{RF as a function of AOD, yearly means. Data from the five simulations listed in
    table~\ref{tab:simulation-overview} (C2W\(\downarrow\), C2W\(-\), C2W\(\uparrow\),
    C2WN\(\uparrow\), and C2W\(\uparrow\uparrow\)) are shown along with the data from the
    HadCM3 sstPiHistVol simulation by \citeA{gregory2016} (grey crosses, G16). Also shown
    are the estimated peak values of the Mt.\ Pinatubo (black star) and Mt.\ Tambora (black
    plus) eruptions. The peak values from the C2W simulations are shown as red circles.
    Additionally in (a) the simulated super-volcano of \citeA{jones2005} (pink square) is
    shown. All peak values (as opposed to annual means) have an asterisk (\(\ast{}\)) in
    their label. The grey lines are the same regression fits as in \citeA<>[Fig.\
      4]{gregory2016}, where the solid line is the fit to G16. (b): Zooming in on the smallest
    AOD values.}\label{fig:2_rf_vs_aod_slopes}%
\end{figure}

We next focus on the development of the AOD and RF time series relative to each other.
Similar comparisons were conducted in \citeA<>[their Fig.\ 4]{gregory2016} and
\citeA<>[their Fig.\ 1]{marshall2020}, with RF plotted against AOD.
Figure~\ref{fig:2_rf_vs_aod_slopes} displays annual mean values from the five simulation
cases in table~\ref{tab:simulation-overview}; the small eruption case
(C2W\(\downarrow\)) as blue downward-pointing triangles, the intermediate eruption case
(C2W\(-\)) as orange thick diamonds, the large tropical eruption case (C2W\(\uparrow\))
as green upward-pointing triangles, the extra large eruption case
(C2W\(\uparrow\uparrow\)) as small pink upward-pointing carets, and the large northern
hemisphere eruption case (C2WN\(\uparrow\)) as brown upward-pointing three-branched
twigs. Also shown are the data from \citeA<>[Fig.\ 4, black crosses from HadCM3
  sstPiHistVol]{gregory2016} as grey crosses labelled G16 (described in Appendix B,
section~\ref{ap:g16}). Additionally, the estimated peak values from the Mt.\ Pinatubo
and Mt.\ Tambora eruptions are plotted as a black star and plus, while the peak from the
\citeA{jones2005} simulation is shown as a pink square labelled J05. Finally, red
circles represent the peak values obtained from the C2W eruption cases. The straight
lines are the same as shown by \citeA{gregory2016}. The full data range is shown in
Fig.~\ref{fig:2_rf_vs_aod_slopes}a while Fig.~\ref{fig:2_rf_vs_aod_slopes}b highlights a
narrow range, focusing on the C2W\(\downarrow\) case.

The annual mean data from the Pinatubo-like C2W\(\downarrow\) case in
Fig.~\ref{fig:2_rf_vs_aod_slopes}b have RF values as a function of AOD that follow
almost the same constant slope as the G16 data. However, in
Fig.~\ref{fig:2_rf_vs_aod_slopes}a we observe that the stronger eruptions lead to
dissimilar responses in AOD and RF, where C2W\(-\) seems to follow close to a \(-10\)
slope and C2W\(\uparrow\) is closer to a \(-5\) slope. The peak values (red circles)
suggest a non-linear dependence, while within each eruption strength (same colour) the
annual mean values fall relatively close to a straight line.

\begin{figure}
  \centering
  \includegraphics{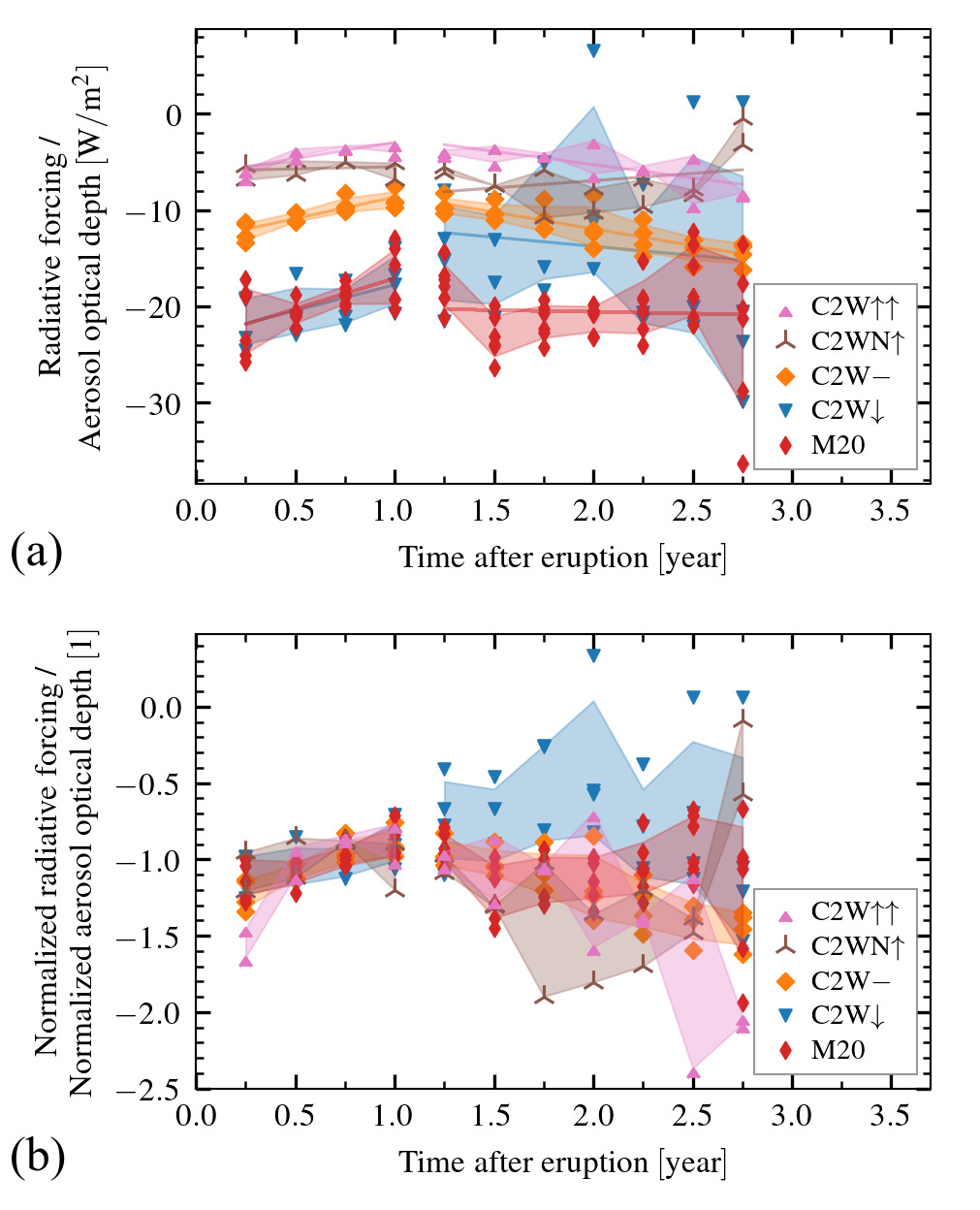}

  \caption{(a): The ratio of RF to AOD, with time-after-eruption on the horizontal axis.
    Straight lines indicate linear regression fits and are described in
    table~\ref{tab:slope-gradients}, while shaded regions are the standard deviation across
    the ensembles for each season. Regression fits and shadings are made for the pre-peak
    and post-peak periods. (b): Same as in (a), but where the underlying AOD and RF time
    series have been scaled to have peak values at unity. Shown are data from
    table~\ref{tab:simulation-overview} along with tropical eruptions from
    M20.}\label{fig:3_rf_to_aod_ratios}%
\end{figure}

To investigate the time dependence of the ratio between RF and AOD, we present seasonal
means of this ratio in Fig.~\ref{fig:3_rf_to_aod_ratios}. The plot shows the eruption
cases given in table~\ref{tab:simulation-overview}, as well as the tropical eruptions
from \citeA{marshall2020dataset} (\(6\) of \(82\) eruptions), labelled M20 and described
in Appendix B, section~\ref{ap:m20}. The C2W\(\uparrow\) case is similar to
C2W\(\uparrow\uparrow\) as indicated in table~\ref{tab:slope-gradients}, but is not
shown in the plot to better highlight C2WN\(\uparrow\). In
Fig.~\ref{fig:3_rf_to_aod_ratios}a, lines are linear regression fits to the seasonal
means across all ensemble members, summarised in table~\ref{tab:slope-gradients}. Shaded
regions are the standard deviation around the seasonal means. A similar shading is
plotted in Fig.~\ref{fig:3_rf_to_aod_ratios}b, but where the regression fits have been
omitted for clarity. As the AOD and RF time series start from zero, the ratio from the
first season is not included. Likewise, after three years both time series are almost
fully equilibrated (Fig.~\ref{fig:1_compare_waveform}a,b). The data is further divided
into two periods; a pre-peak period where the peak of both the AOD and the RF is
included (consisting of the first post-eruption year), and a post-peak period for the
decaying part (consisting of the second and third post-eruption years).

Although the ratio changes across the eruption magnitudes, we find that all the tropical
cases follow a positive slope during the pre-peak period, as seen in
Fig.~\ref{fig:3_rf_to_aod_ratios}a and described in table~\ref{tab:slope-gradients}. The
northern latitude case in C2WN\(\uparrow\) shows a much flatter slope compared to
C2WTrop and M20. The distinction between the slopes from the tropical and non-tropical
cases is perhaps more clear in Fig.~\ref{fig:3_rf_to_aod_ratios}b and corresponding rows
in table~\ref{tab:slope-gradients}. Again, C2WN\(\uparrow\) shows an almost flat slope
compared to the tropical cases. During the post-peak period, more noise is introduced,
but a weak tendency of negative slopes is found among the tropical cases, as well as in
the C2WN\(\uparrow\) case up to the last season where the noise is also the largest.

\begin{table}
  \centering

  \caption{Slope and standard deviation for the data in
    Fig.~\ref{fig:3_rf_to_aod_ratios}\(^{a}\)}\label{tab:slope-gradients}%
  \begin{tabular}{cccc}
    \toprule
    Figure                                         & Ensemble name           & Pre-peak        & Post-peak        \\
    \midrule
                                                   & C2WN\(\uparrow\)        & \(0.45\pm1.15\) & \(1.51\pm1.45\)  \\
                                                   & C2W\(\uparrow\uparrow\) & \(3.38\pm0.97\) & \(-2.74\pm0.77\) \\
    \multirow{2}{*}{\ref{fig:3_rf_to_aod_ratios}a} & C2W\(\uparrow\)         & \(3.85\pm0.52\) & \(-3.29\pm0.60\) \\
                                                   & C2W\(-\)                & \(4.36\pm0.82\) & \(-3.37\pm0.59\) \\
                                                   & C2W\(\downarrow\)       & \(3.64\pm2.41\) & \(-1.41\pm3.25\) \\
                                                   & M20                     & \(6.34\pm1.77\) & \(-0.36\pm1.33\) \\
    \midrule
                                                   & C2WN\(\uparrow\)        & \(0.08\pm0.20\) & \(0.27\pm0.26\)  \\
                                                   & C2W\(\uparrow\uparrow\) & \(0.86\pm0.25\) & \(-0.70\pm0.19\) \\
    \multirow{2}{*}{\ref{fig:3_rf_to_aod_ratios}b} & C2W\(\uparrow\)         & \(0.75\pm0.10\) & \(-0.64\pm0.12\) \\
                                                   & C2W\(-\)                & \(0.43\pm0.08\) & \(-0.34\pm0.06\) \\
                                                   & C2W\(\downarrow\)       & \(0.18\pm0.12\) & \(-0.07\pm0.16\) \\
                                                   & M20                     & \(0.33\pm0.07\) & \(-0.02\pm0.08\) \\
    \toprule
    \multicolumn{4}{l}{\parbox{\linewidth}{\(^{a}\)The regression fits in the top half of the table are for
        Fig.~\ref{fig:3_rf_to_aod_ratios}a, while the bottom half is for
        Fig.~\ref{fig:3_rf_to_aod_ratios}b. The columns ``pre-peak'' and ``post-peak'' refer to
        the two periods as shown in Fig.~\ref{fig:3_rf_to_aod_ratios}. The ensembles are the
        same as those given in table~\ref{tab:simulation-overview}, in addition to the \(6\)
        tropical eruptions from the \(82\) member ensemble in
    \citeA{marshall2020}.}}                                                                                       \\
  \end{tabular}
\end{table}

\citeA<>[their Fig.\ 1c,d]{marshall2020} present results that demonstrate a
time-dependent relationship in the conversion between AOD and RF. They obtain an RF to
AOD ratio with a negative slope when comparing the first post-eruption year to the
second and third. As such, \citeA{marshall2020} find that, on average, the aerosol
forcing efficiency increases during the first two to three post-eruption years. This
phenomenon is explained by \citeA{marshall2020} as the aerosols initially being
spatially confined to the hemisphere where the eruption occurred. Subsequently, during
the second and third years, they spread globally, resulting in a higher global-mean
albedo per AOD and consequently a stronger RF per AOD ratio with time. However, as noted
above, a decrease in aerosol forcing efficiency is found when analysing the M20 data
with seasonal resolution during the pre-peak period (first year post-eruption) while
constraining the ensemble to only include eruptions within \(-10\) to
\(\SI{10}{\degree\mathrm{N}}\). The post-peak period shows an increasing aerosol forcing
efficiency, and during the full first three post-eruption years (pre-peak and
post-peak), both the tropical subset and the full M20 data yield an increasing
efficiency, as expected. Likewise, the first three post-eruption years of the C2W\(-\),
C2W\(\uparrow\uparrow\), and C2WN\(\uparrow\) cases show a weak negative slope and thus
an increasing efficiency, while C2W\(\downarrow\) shows an elevated post-peak ratio as
seen in Fig.~\ref{fig:3_rf_to_aod_ratios}b.

We also note that while the aerosol forcing efficiency is decreasing for tropical M20
data in the pre-peak period, the full dataset shows increasing efficiency. This is in
line with what we find from C2WN\(\uparrow\), which is the only eruption case that does
not show a clear aerosol forcing efficiency decrease during the pre-peak period.

\subsection{Parameter scan}

\begin{figure*}
  \centering
  \includegraphics[width=\textwidth]{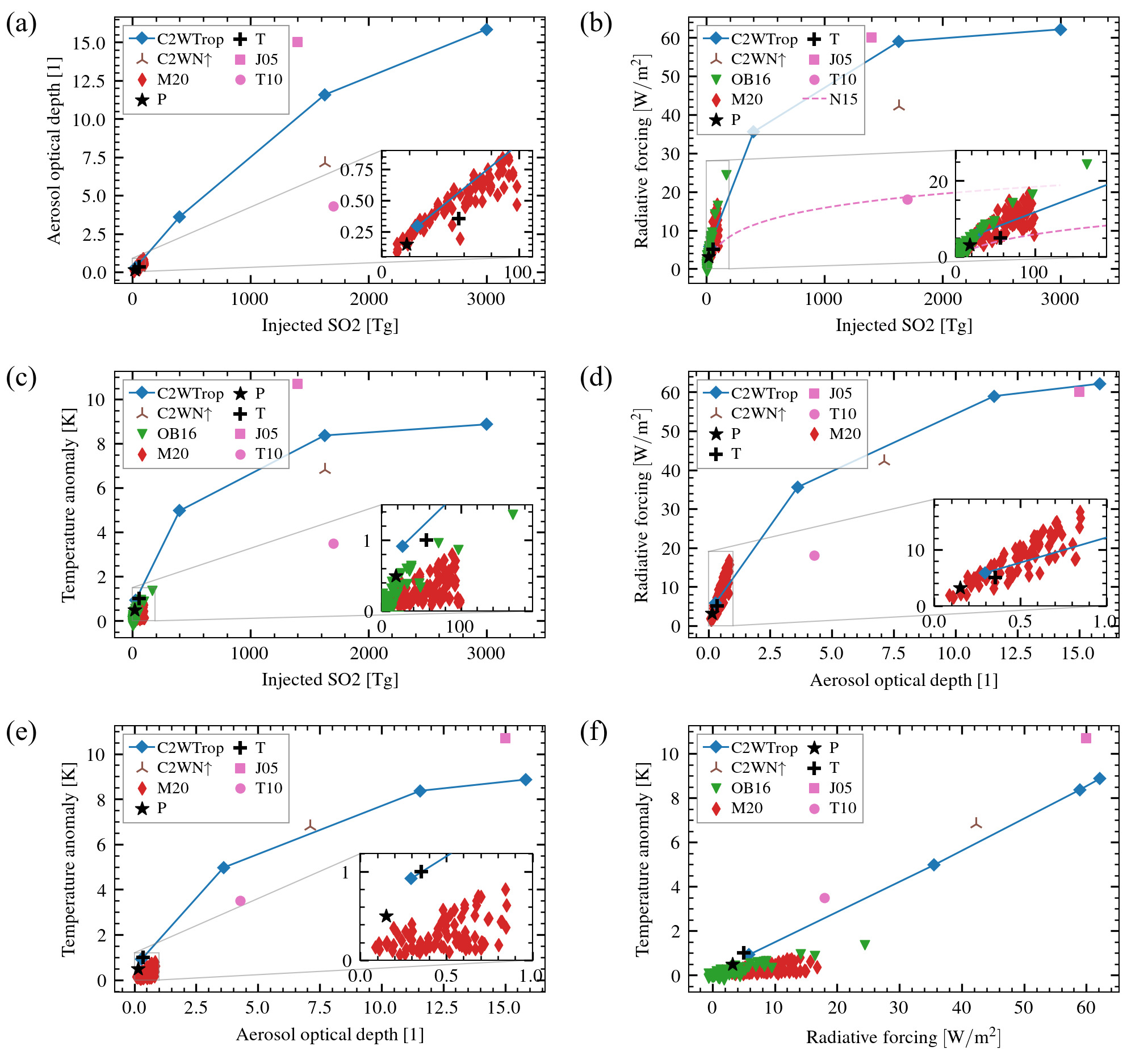}

  \caption{(a) AOD, (b) RF, and (c) temperature anomaly as a function of injected
    \ce{SO2}\@. (d) RF and (e) temperature anomaly as a function of AOD. (f) Temperature
    anomaly as a function of RF. Blue diamonds labelled C2WTrop represent tropical cases
    (C2W\(\downarrow\), C2W\(-\), C2W\(\uparrow\), C2W\(\uparrow\uparrow\)), the brown
    three-branched twig signifies the C2WN\(\uparrow\) case, and green downward triangles
    denote OB16 data from \citeA{ottobliesner2016}. The red thin diamonds labelled M20
    display the \citeA{marshall2020dataset} data. Black star and plus indicate Mt.\ Pinatubo
    and Mt.\ Tambora estimates based on observations. The pink square labelled J05 refers to
    the one-hundred times Mt.\ Pinatubo super-volcano from \citeA{jones2005}, and the pink
    disk labelled T10 represents the YTT super-volcano from \citeA{timmreck2010}. The pink
    dashed line labelled N15 is from \citeA{niemeier2015}, indicating the function in
    Eq.~\ref{eq:niemeier_exponential}.}\label{fig:4_parameter_scan}%
\end{figure*}

In Fig.~\ref{fig:4_parameter_scan}, we compare the peak values of all investigated CESM2
output parameters against each other as well as to injected \ce{SO2}. For our tropical
cases (C2WTrop), we observe in Fig.~\ref{fig:4_parameter_scan}a an almost linear yet
notably weakening relationship between AOD peak values and injected \ce{SO2}. The
latitude also plays a role in the magnitude of the AOD perturbation, evident from
C2WN\(\uparrow\). This weak yet notable latitude dependence aligns with findings by
\citeA{marshall2019}, indicating that \(\SI{72}{\percent}\) of the AOD variance can be
attributed to injected \ce{SO2}, while latitude accounts for only \(\SI{16}{\percent}\)
of the variance. Peak values from their data (82 simulations) plotted as red thin
diamonds display a similar pattern, with AOD exhibiting close to linear dependence on
injected \ce{SO2}, but with latitude introducing a spread in AOD. Peak values from Mt.\
Pinatubo (P) and Mt.\ Tambora (T) are shown for reference, along with peak values from
\citeA{jones2005} labelled J05 and \citeA{timmreck2010} labelled T10. The J05 is a
simulation of a super-volcano based on a \(100\) times scaling of the AOD from Mt.\
Pinatubo, while T10 is a simulation of the YTT eruption based on \ce{SO2} injections.

In Fig.~\ref{fig:4_parameter_scan}b, RF plotted against injected \ce{SO2} (with the
absolute value of RF on the \(y\)-axis) indicates a substantial damping effect on RF as
injected \ce{SO2} increases for the C2W data, in agreement with results from
\citeA{ottobliesner2016}, labelled OB16. The OB16 data come from a \(2500\) year long
simulation using historic volcanoes as the only external forcing. The analysis details
of OB16 can be found in Appendix B, section~\ref{ap:ob16}. Despite the model complexity
difference, \citeA{ottobliesner2016}'s simulations using Community Earth System Model
version 1 (CESM1) with a low-top atmosphere (CAM5) produce RFs comparable to our
findings.

% INFO: the conversion between S and SO2 is confirmed by Niemeier and Timmreck (2015)'s
% reference to the Bekki et al. (1996) paper. Bekki uses 6000 Mt SO2, Niemeier uses 3000
% Tg(S).
\citeA{niemeier2015} conducted simulations of continuous sulphur injections up to
\(\SI{200}{\tera\gram(\ce{SO2})\mathrm{yr}^{-1}}\) in the ECHAM5's middle atmosphere
version \cite{giorgetta2006} with aerosol microphysics from HAM \cite{stier2005}. They
observed an RF dependence on \ce{SO2} injection rate following an inverse exponential,
which converges to \(\SI{-65}{\watt\meter^{-2}}\), depicted in
Fig.~\ref{fig:4_parameter_scan}b as the stippled pink line labelled N15 and given as;

\begin{equation}
  \Delta
  R_{\mathrm{TOA}} =
  -\SI{65}{\watt\metre^{-2}}
  \mathrm{e}^{-{\left(\frac{\SI{2246}{\tera\gram(S)yr^{-1}}}{x}\right)}^{0.23}}.
  \label{eq:niemeier_exponential}
\end{equation}
Both our simulations and OB16 exhibit a notably faster increase than this exponential
relationship. The results by N15, on which Eq.~\ref{eq:niemeier_exponential} is based,
are all averages over at least three years of steady sulphur burdens, substantially
longer than the time it takes for RF to reach peak values after an eruption. Combined
with their lack of a full chemistry model \cite{niemeier2015}, a direct comparison
between Eq.~\ref{eq:niemeier_exponential} to peak RF values (occurring about one year
post-eruption) may not reflect the same chemical and physical processes. In
Eq.~\ref{eq:niemeier_exponential}, \(x\) represents \ce{S}, while the axis shows values
of \ce{SO2}, thus halving of the \ce{SO2} values on the axis gives the appropriate shape
of Eq.~\ref{eq:niemeier_exponential} as a function of \ce{S}.

With these caveats in mind, we observe that T10's results closely align with the
function described in Eq.~\ref{eq:niemeier_exponential}. Starting with an initial input
of \(\SI{850}{\tera\gram(\ce{S})}\) (equivalent to \(\SI{1700}{\tera\gram(\ce{SO2})}\),
representing the YTT eruption), their estimated AOD led to a peak RF of
\(\SI{-18}{\watt\meter^{-2}}\), depicted as a pink filled circle in
Fig.~\ref{fig:4_parameter_scan}b. The results from T10 came from a simulation using the
MPI-ESM climate model, driven by AOD data from the HAM aerosol model. This alignment
likely stems from the utilization of the same aerosol microphysical model in both
\citeA{timmreck2010} and \citeA{niemeier2015}, as well as the application of similar
climate models, MPI-ESM and ECHAM5, respectively. The relationship between climate model
families and their implications are further described in Appendix C. Notably, the peak
values from M20 fit well within an upper boundary defined by C2WTrop and OB16, and a
lower boundary defined by Eq.~\ref{eq:niemeier_exponential}. Eruptions closer to the
equator within M20 align with data points near the upper boundary, whereas eruptions at
more extreme latitudes tend to yield weaker peak RF values, closer to the lower
boundary. Importantly, none of the eruption simulations shown in
Fig.~\ref{fig:4_parameter_scan}b exceeded the upper threshold of
\(\SI{-65}{\watt\meter^{-2}}\) as suggested in Eq.~\ref{eq:niemeier_exponential}.

Figure~\ref{fig:4_parameter_scan}c illustrates the response of temperature against
injected \ce{SO2}. The increase in temperature response with injected \ce{SO2} decreases
for higher injected \ce{SO2}, showing a similar relationship between C2WTrop,
C2WN\(\uparrow\), and OB16. Notably, T10 and J05 exhibit respectively much weaker and
much stronger temperature responses to injected \ce{SO2} than C2WTrop. T10 has a maximum
temperature anomaly of only \(\SI{-3.5}{\kelvin}\) for their
\(\SI{1700}{\tera\gram(\ce{SO2})}\) eruption, while J05 records a substantially larger
maximum temperature anomaly of \(\SI{-10.7}{\kelvin}\). Since the M20 experiment was
conducted with prescribed sea-surface temperatures \cite{marshall2020}, preventing the
temperature from being fully perturbed, we do not focus on the M20 data in the
temperature plots but include them for completeness.

In Fig.~\ref{fig:4_parameter_scan}d, we revisit the relationship between RF and AOD,
focusing on peak values rather than annual and seasonal averages. As previously
discussed, the RF to AOD ratio displays weaker slopes than previous studies, with the
C2W peak values not conforming to a linear trend. The relationship between RF and AOD
suggests potential substantial dependencies on the model and its input parameters, such
as latitude, but most notably to an inherent non-linear RF dependence on AOD. Both the
G16 data in Fig.~\ref{fig:2_rf_vs_aod_slopes} and the J05 data originate from the same
climate model. Similarly to what we find from the C2W data, the ratio is much stronger
for small eruptions in the industrial era (G16) compared to the super-volcano eruption
(J05).

In Fig.~\ref{fig:4_parameter_scan}e, we again find that the response of the C2WTrop data
decreases with injected \ce{SO2}, this time in temperature anomaly. Additionally, both
the C2WN\(\uparrow\) and the J05 cases align well with C2WTrop, with the T10 case
following a similar dependence.

Finally, in Fig.~\ref{fig:4_parameter_scan}f, we compare the temperature and RF
responses. Both C2WTrop and OB16 show a near-linear relationship between temperature and
RF. The C2WTrop data indicate a steeper slope, implying stronger temperature
perturbations as compared to OB16. However, potential biases exist in the values from
the analysis of OB16, as outlined in Appendix B, section~\ref{ap:ob16}. This, along with
considerable noise, results in the analysis of OB16 temperature anomalies being less
reliable. As in Fig.~\ref{fig:4_parameter_scan}e, the C2WN\(\uparrow\) case along with
both the T10 and J05 cases closely follow the temperature to RF dependence of C2WTrop.

The almost linear relationship between AOD and injected \ce{SO2} for the C2WTrop data in
Fig.~\ref{fig:4_parameter_scan}a suggests a comparable trend for RF versus injected
\ce{SO2} in Fig.~\ref{fig:4_parameter_scan}b, as seen for RF versus AOD in
Fig.~\ref{fig:4_parameter_scan}d. For the same reason, we expect
Fig.~\ref{fig:4_parameter_scan}e to show a similar pattern for C2WTrop as observed in
Fig.~\ref{fig:4_parameter_scan}c.

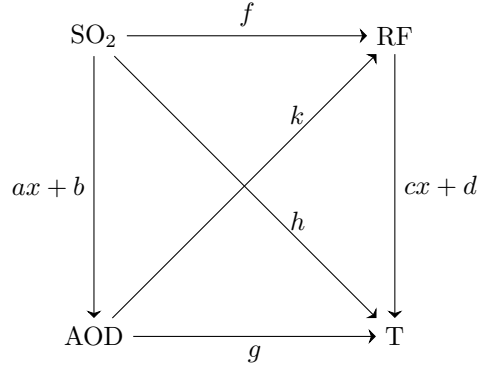
\begin{figure}
  \centering

  \begin{tikzpicture}[>={Stealth[length=1mm,width=2mm]}]
    % Place the letters in the corners
    \node (A) at (0,0) {AOD}; \node (T) at (4,0) {T}; \node (RF) at (4,4) {RF}; \node (S) at
    (0,4) {\ce{SO2}};
    % Draw arrows between the letters
    \draw[->] (A) -- node[midway, below] {\(g\)} (T); \draw[->] (RF) -- node[midway, right]
    {\(cx+d\)} (T); \draw[->] (S) -- node[midway, above] {\(f\)} (RF); \draw[->] (S) --
    node[midway, left] {\(ax+b\)} (A); \draw[->] (S) -- node[pos=0.7, above] {\(h\)} (T);
    \draw[->] (A) -- node[pos=0.7, above] {\(k\)} (RF);
  \end{tikzpicture}

  \caption{Diagram describing the functional relationships of the parameters shown in
    Fig.~\ref{fig:4_parameter_scan}.}\label{fig:5_diagram_of_function_relations}%
\end{figure}

This relationship, along with the functional relationships between all other parameters
shown in Fig.~\ref{fig:4_parameter_scan}, are illustrated in
Fig.~\ref{fig:5_diagram_of_function_relations}. There, we show that from assuming a
linear dependency of AOD on injected \ce{SO2} (\(ax+b\)), and of temperature on RF
(\(cx+d\)), we must have that \(f\), \(g\), \(h\), and \(k\) all have the same
functional form, where \(f: \ce{SO2} \to \mathrm{RF}\), \(g: \mathrm{AOD} \to
\mathrm{T}\), \(h: \ce{SO2} \to \mathrm{T}\), and \(k: \mathrm{AOD} \to \mathrm{RF}\).
From this, we deduce that \(f(x)=k(ax+b)\) and \(h(x)=f(cx+d)=g(ax+b)\), and finally
that \(h(x)=k(acx+ad+b)\), concluding that \(f\), \(g\), \(h\), and \(k\) have the same
functional form.

\subsection{Climate sensitivity estimate}

As previously mentioned, the J05 experiment is similar to C2W\(\uparrow\) concerning RF
values, yet differ in both AOD and temperature. At the same time J05 is similar to
C2W\(\uparrow\uparrow\) in AOD and RF. To investigate this discrepancy, we here conduct
a comparison between the J05 climate feedback parameter \(\alpha\) (where \(s=1/\alpha\)
is the climate sensitivity parameter) with our climate resistance, denoted as \(\rho\),
and the transient climate response parameter (TCRP) \(1/\rho\) (where
\(\mathrm{TCS}=F_{2\times\ce{CO2}}\times \mathrm{TCRP}\) is the transient climate
sensitivity and \(F_{2\times\ce{CO2}}\) is the forcing due to a doubling of
pre-industrial \ce{CO2} concentration). As the forcing of volcanic eruptions typically
lasts for about a year, a duration too short for \(F=\rho T\) to remain valid, an
alternative approach using a time-integral form introduced by \citeA{merlis2014} is
applied:

\begin{equation}
  \int_0^{\tau}F \mathrm{d}t=\rho\int_{0}^{\tau}T \mathrm{d}t
  \label{eq:climate-resistance-orig}
\end{equation}
\begin{equation}
  \rho=\frac{\int_0^{\tau}F \mathrm{d}t}{\int_{0}^{\tau}T \mathrm{d}t}.
  \label{eq:climate-resistance}
\end{equation}

If the upper bound of the integral, \(\tau\), is sufficiently large, so that the upper
ocean heat content is the same at \(t=0\) and \(t=\tau\) (\citeA{merlis2014} used \(\tau
=\SI{15}{\mathrm{yr}}\)), this approach agrees with \(F=\rho T\) for long-term forcing
\cite{gregory2016}. Additionally, we note that the climate resistance and the climate
feedback parameter are associated with the ocean heat uptake efficiency (\(\kappa\))
through \(\rho =\alpha +\kappa\) \cite{gregory2016}.

The climate feedback parameter estimated by \citeA{jones2005} is \(\alpha \simeq
\SI{4}{\watt\metre^{-2}\kelvin^{-1}}\), exceeding twice the value obtained by
\citeA{gregory2016} in their simulations of Mt.\ Pinatubo using the same HadCM3 climate
model. We determine the climate resistance using the integral-form computation outlined
in Eq.~\ref{eq:climate-resistance} and adopting \(\tau =\SI{20}{\mathrm{yr}}\). The
estimated climate resistance from the three tropical simulation cases (with four in each
ensemble) converges to \(\rho =\SI{2.5(4)}{\watt\metre^{-2}\kelvin^{-1}}\), and TCRP
values of \(1/\rho=\SI{0.41(5)}{\kelvin\watt^{-1}\metre^{2}}\), as reported in
table~\ref{tab:trcp}, and is therefore assumed to be a good estimate of \(\alpha\).

Importantly, our estimate agrees well with G16, while the J05 estimate of \(\alpha
\simeq \SI{4}{\watt\metre^{-2}\kelvin^{-1}}\) is still notably higher. Since the
temperature perturbation obtained by J05 was larger than in any of our CESM2 cases, it
indicates that the forcing used by J05 must be stronger. The peak value of the J05 RF is
similar to the C2W\(\uparrow\) case, and as such, the overall stronger forcing must
originate from the development of the forcing time series rather than the peak value.

\begin{table}
  \centering

  \caption{Estimated climate resistance and TCRP\(^{a}\)}\label{tab:trcp}%
  \begin{tabular}{ccc}
    \toprule
    Simulation type   & \(\rho [\si{\watt\metre^{-2}\kelvin^{-1}}]\) & \(1/\rho\)            \\
    \midrule
    C2W\(\uparrow\)   & \(\num{2.21\pm0.05}\)                        & \(\num{0.45\pm0.01}\) \\
    C2W\(-\)          & \(\num{2.51\pm0.06}\)                        & \(\num{0.40\pm0.01}\) \\
    C2W\(\downarrow\) & \(\num{2.9\pm0.6}\)                          & \(\num{0.36\pm0.07}\) \\
    Total             & \(\num{2.5\pm0.4}\)                          & \(\num{0.41\pm0.05}\) \\
    \toprule
    \multicolumn{3}{l}{\parbox{\linewidth}{\(^{a}\)Estimates are based on ensembles with four members and \(\tau
    =\SI{20}{\mathrm{yr}}\) using Eq.~\ref{eq:climate-resistance}.}}                         \\
  \end{tabular}
\end{table}

\section{Discussion}\label{sec:discussion}

% NOTE: Suggested layout for the
% Discussion:
% - Explain the results and emphasize significant findings clearly
% - Discuss the impact and importance of results compared with recent relevant research
% Conclusion
% - The justification for these objectives: Why is the work important?
% - Summarize the key points made in the other sections
% - Conclude overall discussion of article
% - Link this section to the introduction

% RF / AOD linearity does not hold for large eruptions
Figures~\ref{fig:2_rf_vs_aod_slopes},~\ref{fig:3_rf_to_aod_ratios}, and
\ref{fig:4_parameter_scan}d demonstrate that as the AOD exceeds approximately \(1.0\),
the linear RF dependence of approximately
\(\SI{-20}{\watt\metre^{-2}\mathrm{AOD}^{-1}}\) no longer holds. The sublinear increase
in RF with injected \ce{SO2} in Fig.~\ref{fig:4_parameter_scan}b for large eruptions is
consistent with previous results from simulations using similar climate models of
smaller historic eruptions (G16) and of super-volcanoes (J05). Such a change in ratio
has been attributed to larger eruptions, injecting more \ce{SO2}, leading to larger
aerosols, and hence less effective radiation scattering, thereby reducing the RF for the
same injected \ce{SO2} \cite{english2013, timmreck2010, timmreck2018}.

% Latitude is more important for the development of the RF / AOD ratio
The non-linear relationship between peak RF and AOD values is a strong signature in both
Figs.~\ref{fig:2_rf_vs_aod_slopes} and~\ref{fig:3_rf_to_aod_ratios}. Across eruptions of
the same strength, the ratio stays relatively constant, leading to a close to
\(\SI{-10}{\watt\metre^{-2}\mathrm{AOD}^{-1}}\) slope for C2W\(-\) and a
\(\SI{-5}{\watt\metre^{-2}\mathrm{AOD}^{-1}}\) slope for C2W\(\uparrow\) and
C2W\(\uparrow\uparrow\). Still, a non-linear development in the RF to AOD ratio is found
across all tropical eruptions. Similar to the results of \citeA{marshall2020}, we find
in C2W\(-\), C2W\(\uparrow\), and C2W\(\uparrow\uparrow\) that the post-peak period
(second and third post-eruption years) has a stronger aerosol forcing efficiency
compared to the pre-peak period (first post-eruption year). The post-peak period of
C2W\(\downarrow\) is elevated as compared to the pre-peak period, resulting in a
decreasing aerosol forcing efficiency from the first to the second and third
post-eruption years, in contrast to the other tropical eruptions.

Focusing on the pre-peak period, we find tropical eruptions to differ from eruptions at
high latitudes. During the pre-peak period, all tropical eruptions show a decreasing
aerosol forcing efficiency, while no significant change in the RF to AOD ratio is found
from the C2WN\(\uparrow\) case. The full M20 dataset indicates an increasing aerosol
forcing efficiency also during the pre-peak period, contrasting the decreasing
efficiency found from their tropical eruptions and supporting the latitudinal dependence
we find with C2WN\(\uparrow\). While we find a linear relationship to be a useful
approximation of RF dependence on AOD for eruptions similar to or smaller than Mt.\
Pinatubo, additional factors must be considered for larger eruptions. These factors,
such as \ce{OH} scarcity and aerosol growth, influence reflectance and their
gravitational pull, substantially impacting both AOD and RF evolution, is highlighted by
\citeA{timmreck2010}. The large difference in ratio found when comparing eruption
magnitudes suggests that injected \ce{SO2} is crucial when estimating the time-average
of the RF to AOD ratio. However, latitude and, in particular, aerosol dispersion are
more influential in determining the post-eruption evolution of the ratio, particularly
during the pre-peak period.

We find that the suggested upper threshold from Eq.~\ref{eq:niemeier_exponential} is not
violated by any eruption simulation, and most notably that the temperature peak value
follow the RF trend in reaching a limiting value. The C2WTrop cases follow a close to
linear temperature dependency on RF, with the J05, T10, and C2WN\(\uparrow\) aligning
close to the same slope. The linear relationship between temperature and RF is the
strongest dependence found between the parameters in Fig.~\ref{fig:4_parameter_scan},
and a strong signature across both eruption magnitudes and latitudes, but also across
highly different climate models. Thus, from a maximum RF of
\(\SI{-65}{\watt\metre^{-2}}\), we expect temperature anomalies to reach at most
\(\sim\SI{-12}{\kelvin}\).

% Models vary in their SO2 to H2SO4 chemistry/physics, but are comparatively consistent in representing AOD, RF and temperature.
The biggest spread in the data shown in Fig.~\ref{fig:4_parameter_scan} is found when
relating injected \ce{SO2} to any of the three output parameters. As the amount of
injected \ce{SO2} increases, both AOD, RF, and temperature across models have a big
spread. The AOD to injected \ce{SO2} relationship is consistent within similar models,
even when comparing simulations of volcanic eruptions \cite{timmreck2010} and continuous
injection of \ce{SO2} \cite{niemeier2015}, but has a wide spread at high values of
injected \ce{SO2} across model families (Figs.~\ref{fig:4_parameter_scan}a,b,c).
Comparatively, the RF (Fig.~\ref{fig:4_parameter_scan}d) and temperature
(Fig.~\ref{fig:4_parameter_scan}e) as a function of AOD, as well as temperature as a
function of RF (Fig.~\ref{fig:4_parameter_scan}f), demonstrate a smaller spread across
models. \citeA{marshall2019, marshall2020, marshall2021} use a code with seven
log-normal modes to simulate aerosol mass and number concentrations, along with an
atmosphere-only configuration of the UM-UKCA with prescribed sea-surface temperatures
and sea-ice extent \cite{marshall2019}. This approach is in contrast with CESM2,
operating as an Earth System Model, but with a simpler aerosol chemistry model in the
MAM3. The family of models to which M20 is based is different from that of C2W and OB16,
and also different from the T10 and N15, as described in Appendix C. Based on
Fig.~\ref{fig:4_parameter_scan}, we find the model family to be pivotal in determining
the estimated AOD and RF magnitudes from injected \ce{SO2}, whereas the various models
generally demonstrate more consistency in representing RF from AOD.

% AOD time series are different across eruption magnitudes, and peak at later times
\citeA{timmreck2010} highlights that for sufficiently large eruptions, \ce{OH} radicals
are too scarce, which limits \ce{SO2} oxidation. The AOD peak in the YTT simulation of
T10 occurs six months after Mt.\ Pinatubo's peak. This aligns with our results, as
illustrated in Fig.~\ref{fig:1_compare_waveform}a, where C2W\(\downarrow\) shows an
earlier AOD peak compared to C2W\(-\), C2W\(\uparrow\), and C2W\(\uparrow\uparrow\).
While the peak RF value of T10 occurs \(7\)--\(8\) months post-eruption, similar to C2W,
the J05 peak anomaly occurs one year post-eruption. Additionally, as \citeA{jones2005}
obtains a climate feedback parameter larger than both what \citeA{gregory2016} found for
the same climate model and larger than the climate resistance obtained here from C2W, we
conclude that such a simple approach of scaling the AOD of smaller eruptions to
represent larger eruptions is insufficient. Moreover, having a small ensemble of large
eruptions to represent smaller eruptions is also insufficient when simulating from
injected \ce{SO2}, as both AOD and temperature evolution are found to develop
differently.

\section{Summary and conclusions}\label{sec:conclusions}

We consider five medium to super-volcano sized eruption ensembles and compare them to
previously reported results. We find the commonly adopted RF dependence on AOD of
\(\sim\SI{-20}{\watt\metre^{-2}\mathrm{AOD}^{-1}}\) to be representative for Mt.\
Pinatubo-sized eruptions. Larger eruptions, with one to two orders of magnitude larger
injections of \ce{SO2}, are found to have an RF dependence on AOD closer to
\(\sim\SI{-10}{\watt\metre^{-2}\mathrm{AOD}^{-1}}\) and
\(\sim\SI{-5}{\watt\metre^{-2}\mathrm{AOD}^{-1}}\). A shallower slope for larger
eruptions is also consistent with peak values from previous studies of super-volcanoes.

The time-after-eruption dependence of the ratio between RF and AOD is found to weaken
with time, resulting in a decreasing aerosol forcing efficiency in the pre-peak period.
The effect is found across all eruption sizes, but only the tropical cases show a clear
trend. The high-latitude case displays an almost constant efficiency with time. These
results agree with a reanalysis of the tropical data in \citeA{marshall2020dataset}.
Thus, these findings provide strong supporting evidence that latitude is generally
significant in determining the aerosol forcing efficiency, particularly as a function of
time-after-eruption. These findings emphasise the complexity of volcanic impacts on
climate, demonstrating significant differences in climatic response depending on
eruption magnitude and latitude.

We find that the AOD peak arrives later for larger eruptions than for smaller ones, and
also that larger eruptions produce a sharper peak in the AOD time series. The RF time
series are similar across all eruption sizes, and while the smallest eruption
experiences a faster temperature decay, the larger eruptions produce time series
indistinguishable in development for both RF and temperature. Thus, a simple scaling of
the AOD or temperature time series from a smaller eruption is insufficient in
representing that of larger volcanic eruptions.

Considering injected \ce{SO2} and the peak values of AOD and RF, a large spread is found
across model families in Fig.~\ref{fig:4_parameter_scan}. Improving the consistency
between model families in how the chemistry and physics of \ce{SO2} and \ce{H2SO4} are
represented is an important step in enhancing the accuracy of simulated volcanic
eruptions' influence on climate by models. More simulations of larger volcanic eruptions
with injected \ce{SO2} greater than \(\SI{200}{\tera\gram(\mathrm{SO2})}\) would provide
useful information for a more precise determination of the RF to AOD ratio in the
non-linear regime. This would also serve as a useful test to check if a comparison
between \ce{SO2} injection events and continuous \ce{SO2} injection is reasonable.
Introducing a spread in latitude similar to the \citeA{marshall2020dataset} dataset
would allow for better comparison between eruptions across all latitudes and the
suggested lower limit following Eq.~\ref{eq:niemeier_exponential}, describing a
situation of aerosol saturation.

\appendix
\section{Simulation set up and output}

Input files used in the simulations were created by modifying the file available at
\url{http://svn.code.sf.net/p/codescripts/code/trunk/ncl/emission/createVolcEruptV3.ncl},
using a Python package available on GitHub at
\url{https://github.com/engeir/volcano-cooking} or through the Python Package Index
(PyPI). The package is available both as a library and a Command Line Interface (CLI),
and is used to create volcanic eruptions with a specified amount of \ce{SO2} that is
injected over six hours at a given latitude, longitude, and altitude. All volcanic
\ce{SO2} files are created from a shell script by setting the eruption details in a JSON
file that is read by the \texttt{volcano-cooking} CLI at a fixed version, ensuring a
reproducible experiment setup.

We are using the coupled model version \texttt{BWma1850} component setup to run the
CESM2, and an accompanying fixed sea-surface temperature version, \texttt{fSST1850}, to
obtain estimates of the RF. The applied \texttt{fSST1850} is not from a standardised
component setup but is instead explicitly specified as \url{1850_CAM60%WCCM_CLM50%BGC-CROP_CICE%PRES_DOCN%DOM_MOSART_CISM2%NOEVOLVE_SWAV_TEST
  %
}. % Necessary to make the formatter not break the URL / literal text
The component setup \texttt{BWma1850} and \texttt{fSST1850} differ in that the latter
uses a prescribed sea-ice (\texttt{CICE -> CICE\%PRES}), a prescribed data ocean
(\texttt{POP2\%ECO\%DEP -> DOCN\%DOM}) and a stub wave component instead of the full
Wave Watch version 3 (\texttt{WW3 -> SWAV}).

The important input data used in the model simulations are injected \ce{SO2} in units of
teragrams (\(\si{\tera\gram(\ce{SO2})}\)), used to simulate volcanic eruptions. RF is
calculated as the combined (short wave and long wave) all-sky TOA energy imbalance,
where the CESM2 provide the output variables ``net solar flux at the top of the model''
(FSNT) and ``net longwave flux at the top of the model'' (FLNT). Thus, \(\mathrm{RF_*}=
\mathrm{FSNT} - \mathrm{FLNT}\), and taking the difference between volcanic forcing
simulations and a control simulation gives the final estimate of RF
(\(\mathrm{RF}=\mathrm{RF_{VOLC}}-\mathrm{RF_{CONTROL}}\)) \cite{marshall2020}. The RF
calculation is based on \texttt{fSST1850}, hence this outline specifically describes how
to calculate ERF as opposed to IRF, which instead is the difference between the ERF and
the sum of all rapid atmospheric adjustments \cite{marshall2020,smith2018}. The AOD is
obtained from the output variable ``stratospheric aerosol optical depth 550 nm day
nigth'' (AODVISstdn), while global temperature is saved by CESM2 to the variable
``reference height temperature'' (TREFHT). The analysis of this work is performed using
these four variables.

During analysis, one outlier was found in the ensemble representing C2W\(\downarrow\),
specifically in the temperature time series. This ensemble member was the February 15,
1850, eruption, which was changed in favor of a February 15, 1851, eruption in the
C2W\(\downarrow\), C2W\(-\), and C2W\(\uparrow\) ensembles. For completeness, the
February 15, 1850, eruption is still included in the online archive.

\section{External data}

\subsection{Otto-Bliesner data analysis}\label{ap:ob16}

Data from \citeA{ottobliesner2016} are the original input data of injected \ce{SO2} as
used in their model simulations, along with RF and temperature output data. The injected
\ce{SO2} can be found at
\url{https://www.cesm.ucar.edu/working-groups/paleo/simulations/ccsm4-lm}. Only the peak
values of the \ce{SO2} dataset were used in the analysis. Output variables are available
at \url{www2.cesm.ucar.edu/models/experiments/LME}.

Since the OB16 dataset contains a five-member ensemble, the final RF and temperature
time series used were ensemble means. A single control simulation time series is used to
remove seasonal dependence from the temperature, where the control simulation is
averaged into a climatology mean. Further, a drift in the temperature is removed by
subtracting a linear regression fit. RF has seasonality removed in the Fourier domain.

The time of an eruption is found based on a best attempt at aligning the \ce{SO2} time
series with both the RF time series and the temperature time series. The RF and
temperature peak values are taken as the value of the time series at the time of an
eruption according to the \ce{SO2} time series. Missing the true peak means the found
peaks will be biased towards lower values. However, instances where eruptions occur
close in time will contribute a bias to higher values. These biases contribute to a
greater uncertainty related to OB16 in Figs.~\ref{fig:4_parameter_scan}b,c,f.

\subsection{Marshall data analysis}\label{ap:m20}

Data used to compute the M20 values were from \citeA{marshall2020dataset}, available at
\url{https://doi.org/10.5285/232164e8b1444978a41f2acf8bbbfe91}. As each file includes a
single eruption, peak values of AOD, RF, and temperature were found by applying a
Savitzky-Golay filter of third order and one-year window length, and choosing the
maximum value \cite{savitzky1964}.

\subsection{Gregory data analysis}\label{ap:g16}

Data used to compute G16 values were kindly provided by Jonathan Gregory (personal
communication). The full 160-year-long time series were further analysed by computing
annual means.

\section{Model families}

The model used here was the CESM2 with the WACCM6 atmosphere in the MA configuration.
The MA configuration uses the MAM3 \cite{gettleman2019}, a simplified and
computationally efficient default setting within the CAM5 \cite{liu2016}, as described
in \citeA{liu2012}. The MAM3 was developed from MAM7, consisting of the seven modes
Aitken, accumulation, primary carbon, fine dust, fine sea salt, coarse dust, and coarse
sea salt. Instantaneous internal mixing of primary carbonaceous aerosols with secondary
aerosols and instantaneous ageing of primary carbonaceous particles are assumed by
emitting primary carbon in the accumulation mode \cite{liu2016}. As dust absorbs water
efficiently and is expected to be removed by wet deposition similarly to sea salt, fine
dust is merged with fine sea salt into the accumulation mode and coarse dust is merged
with coarse sea salt into a coarse mode. The coarse mode will quickly revert to its
background state below the tropopause \cite{liu2012}. Consequently, MAM3 features the
three modes Aitken, accumulation, and coarse \cite{liu2016}.

The CESM2 is an ancestor of CESM1 used by OB16. They belong to a different model family
than both the HadCM3 (J05 and G16) and the UM-UKCA (M20), which is an extended version
of HadGEM3 \cite{dhomse2014}, and an ancestor of HadCM3. A third model family is
represented through ECHAM5 (N15) and MPI-ESM (T10), where the latter is related to the
former via the ECHAM6. A summary of the model code genealogy is in
table~\ref{tab:model-family}, based on the model code genealogy map created by
\citeA{kuma2023}.

\begin{table*}
  \centering
  \caption{Model code family relations\(^{a}\)}\label{tab:model-family}

  \begin{tabular}{ccc}
    \toprule
    Family relation                                                         & Model name & Data name \\
    \midrule
    \multirow{2}{*}{CESM1 \(\rightarrow\) CESM1-CAM5 \(\rightarrow\) CESM2} & CESM1      & OB16      \\
                                                                            & CESM2
                                                                            & \emph{This
    contribution}                                                                                    \\
                                                                            & HadCM3
                                                                            & J05, G16               \\
    \multirow{-2}{*}{\shortstack{HadCM3 \(\rightarrow\) HadGEM1
    \(\rightarrow\)                                                                                  \\
    HadGEM2 \(\rightarrow\) HadGEM3 \(\rightarrow\) UM-UKCA}}               & UM-UKCA    &
    M20                                                                                              \\
    \multirow{2}{*}{ECHAM5 \(\rightarrow\) ECHAM6 \(\rightarrow\) MPI-ESM}  & ECHAM5     &
    N15                                                                                              \\
                                                                            & MPI-ESM    & T10       \\
    \toprule
    \multicolumn{3}{l}{\parbox{\linewidth}{\(^{a}\)Overview of various model codes grouped into families according to the model
        code genealogy map by \citeA{kuma2023}, with each table entry also indicating the
        specific model code used in the referenced papers of this
    study.}}                                                                                         \\
  \end{tabular}
\end{table*}

%% NOTE that acronyms in the final published version will be spelled out when used in figure captions.
\begin{acronyms}
  \acro{AODVISstdn} ``stratospheric aerosol optical depth 550 nm day night'' \acro{AOD}
  stratospheric aerosol optical depth \acro{CAM5} Community Atmosphere Model Version 5
  \acro{CESM1} Community Earth System Model Version 1 \acro{CESM2} Community Earth
  System Model Version 2 \acro{ECS} equilibrium climate sensitivity \acro{ERF} effective
  radiative forcing \acro{FLNT} ``net longwave flux at the top of the model''
  \acro{FSNT} ``net solar flux at the top of the model'' \acro{IRF} instantaneous
  radiative forcing \acro{MAM3} three mode version of the Modal Aerosol Module \acro{MA}
  middle atmosphere \acro{POP2} Parallel Ocean Program Version 2 \acro{RF} effective
  radiative forcing \acro{TCRP} transient climate response parameter \acro{TOA}
  top-of-the-atmosphere \acro{TREFHT} ``reference height temperature'' \acro{WACCM6}
  Whole Atmosphere Community Climate Model Version 6 \acro{YTT} Young Toba Tuff
\end{acronyms}

\section*{Open Research Section}

% This section MUST contain a statement that describes where the data supporting the
% conclusions can be obtained. Data cannot be listed as ''Available from authors'' or
% stored solely in supporting information. Citations to archived data should be included
% in your reference list. Wiley will publish it as a separate section on the paper’s page.
% Examples and complete information are here: https://www.agu.org/Publish with
% AGU/Publish/Author Resources/Data for Authors

The direct output data of CESM2 are too large to be easily archived and transferred.
Instead, data generated directly from output fields of CESM2 are made available in a
NIRD Research Data Archive \cite{enger2024}, and were generated using scripts available
at \url{https://github.com/engeir/cesm-data-aggregator}. Analysis scripts are available
at GitHub
(\url{https://github.com/engeir/code-to-radiative-forcing-by-super-volcano-eruptions})
and is published to Zenodo \cite{enger2024analysis}. Source code used to generate CESM2
input files are available at \url{https://github.com/engeir/cesm2-volcano-setup}.

\acknowledgments

The simulations were performed on resources provided by Sigma2 --- the National
Infrastructure for High Performance Computing and Data Storage in Norway.

This work was supported by the Tromsø Research Foundation under Grant Number 19\_SG\_AT.

Thanks to both Maria Rugenstein and Martin Rypdal for valuable discussions. We would
also like to thank the authors of \citeA{gregory2016}, \citeA{ottobliesner2016}, and
\citeA{marshall2020dataset} for making their data available.

\bibliography{references}

\end{document}